\documentclass[10pt, conference]{IEEEtran}
\usepackage{graphics, psfrag, amsmath, amsfonts, amssymb, eucal, epsfig}
\newtheorem{prop}{Proposition}

\DeclareMathAlphabet{\eurm}{U}{eur}{m}{n}
\DeclareMathAlphabet{\mathbsf}{OT1}{cmss}{bx}{n}
\DeclareMathAlphabet{\mathssf}{OT1}{cmss}{m}{sl}
\DeclareMathAlphabet{\mathcsf}{OT1}{cmss}{sbc}{n}

\newcommand{\randomvalue}[1]{\eurm{\uppercase{#1}}}

\DeclareSymbolFont{bsfletters}{OT1}{cmss}{bx}{n}  
\DeclareSymbolFont{ssfletters}{OT1}{cmss}{m}{n}
\DeclareMathSymbol{\bsfGamma}{0}{bsfletters}{'000}
\DeclareMathSymbol{\ssfGamma}{0}{ssfletters}{'000}
\DeclareMathSymbol{\bsfDelta}{0}{bsfletters}{'001}
\DeclareMathSymbol{\ssfDelta}{0}{ssfletters}{'001}
\DeclareMathSymbol{\bsfTheta}{0}{bsfletters}{'002}
\DeclareMathSymbol{\ssfTheta}{0}{ssfletters}{'002}
\DeclareMathSymbol{\bsfLambda}{0}{bsfletters}{'003}
\DeclareMathSymbol{\ssfLambda}{0}{ssfletters}{'003}
\DeclareMathSymbol{\bsfXi}{0}{bsfletters}{'004}
\DeclareMathSymbol{\ssfXi}{0}{ssfletters}{'004}
\DeclareMathSymbol{\bsfPi}{0}{bsfletters}{'005}
\DeclareMathSymbol{\ssfPi}{0}{ssfletters}{'005}
\DeclareMathSymbol{\bsfSigma}{0}{bsfletters}{'006}
\DeclareMathSymbol{\ssfSigma}{0}{ssfletters}{'006}
\DeclareMathSymbol{\bsfUpsilon}{0}{bsfletters}{'007}
\DeclareMathSymbol{\ssfUpsilon}{0}{ssfletters}{'007}
\DeclareMathSymbol{\bsfPhi}{0}{bsfletters}{'010}
\DeclareMathSymbol{\ssfPhi}{0}{ssfletters}{'010}
\DeclareMathSymbol{\bsfPsi}{0}{bsfletters}{'011}
\DeclareMathSymbol{\ssfPsi}{0}{ssfletters}{'011}
\DeclareMathSymbol{\bsfOmega}{0}{bsfletters}{'012}
\DeclareMathSymbol{\ssfOmega}{0}{ssfletters}{'012}

\newcommand{\rvn}{{\randomvalue{n}}}	

\newcommand{\rvs}{{\randomvalue{s}}}	

\newcommand{\rvw}{{\randomvalue{w}}}	

\newcommand{\rvx}{{\randomvalue{x}}}	

\newcommand{\rvy}{{\randomvalue{y}}}	

\begin{document}

\title{A Spectrum-Shaping Perspective on Cognitive Radio}
\author{\authorblockN{Wenyi Zhang}
\authorblockA{Qualcomm Research Center\\
5775 Morehouse Drive\\
San Diego, CA, 92121\\
Email: wenyizha@gmail.com}
\and
\authorblockN{Urbashi Mitra}
\authorblockA{Ming Hsieh Dept. of Electrical Engineering\\
University of Southern California\\
Los Angeles, CA, 90089\\
Email: ubli@usc.edu}
}
\maketitle

\begin{abstract}
A new perspective on cognitive radio is presented, where the pre-existent legacy service is either uncoded or coded and a pair of cognitive transceivers need be appropriately deployed to coexist with the legacy service. The basic idea underlying the new perspective is to exploit the fact that, typically, the legacy channel is not fully loaded by the legacy service, thus leaving a non-negligible margin to accommodate the cognitive transmission. The exploitation of such a load margin is optimized by shaping the spectrum of the transmitted cognitive signal. It is shown that non-trivial coexistence of legacy and cognitive systems is possible even without sharing the legacy message with the cognitive transmitter. Surprisingly, the optimized cognitive transmitter is no longer limited by its interference power at the legacy receiver, and can always transmit at its full available device power. Analytical development and numerical illustration are presented, in particular focusing on the logarithmic growth rate, {\it i.e.}, the prelog coefficient, of cognitive transmission in the high-power regime.
\end{abstract}

\section*{Introduction and Overview}
\label{sec:intro}

Cognitive radio is envisioned as a means of improving utilization of scarce spectral resources, and has been attracting increasing attention from both scholars and practitioners (see, {\em e.g.}, \cite{mitola00:phd, haykin05:jsac, sahai06:report}, and papers archived in \cite{dyspan:proc, jsac07:special, jsac08:special} for an incomplete list of topics and results). Loosely speaking, in cognitive radio systems, unlicensed ``cognitive'' users potentially ``sense'' the spectrum utilization and transmit in such a manner as to cause no ``harmful'' interference for legacy users who have been licensed to use the spectrum.

In the prevailing approach to cognitive radio, cognitive transmitters detect the presence of legacy service and if there is none, they transmit thus utilizing the ``white space''. The key challenge there has been that the legacy service detection problem (a.k.a. ``spectrum sensing'') can be extremely difficult in the low-power regime \cite{sahai06:report}, especially in a dynamic environment where the system parameters exhibit a certain amount of uncertainty. In fact, as theoretically shown and numerically validated in \cite{sahai06:report}, there normally exists a threshold (called the SNR wall) such that, if the legacy signal power falls below that threshold, then reliable detection becomes impossible even if an infinitely long observation is made available. Even without such considerations, this conceptually simple approach of filling spectral holes (also called ``white space'' by some practitioners) in nature limits cognitive transmission to spectrally sparse envorinments, and inevitably achieves limited throughput in spectrally ``dense'' ({\em e.g.} urban) areas.

Primarily motivated by the difficulties and limitations of the prior approach, there has been increased attention towards overlay cognitive radio systems, in which cognitive transmission coexists with legacy transmission through proper interference management. From a theoretical point of view, as long as the legacy service does not communicate at its channel capacity, {\it i.e.}, the legacy channel is not ``fully loaded'', it is able to tolerate a certain amount of interference from the cognitive users. In practice, many legacy systems are broadcast systems with many potential subscribers located within a large geographic area, for example, analog/digital television, terrestrial/satellite radio, and cellular downlink. Since a broadcaster needs to provide a uniform quality of service over a long distance, for those receivers not close to the edge of service area, there exist abundant signal-to-noise (SNR) margins beyond the minimum required SNR for successful reception.

Therefore, by imposing a fixed quality of service measure on legacy receivers, we can proceed to design the signal of cognitive transmission subject to that measure. In this paper, we take the quality of service measure as the mean-squared estimation error (MSE) of the legacy signal, in the case of uncoded legacy transmission, and as the achievable information rate of encoding the legacy message, in the case of coded legacy transmission. In the overlay cognitive radio approach, it is not the absence of, but rather the existence of, legacy transmission that we seek to exploit. Departing from this basic idea, the most natural design philosophy perhaps is that, a cognitive transmitter is constrained in a way such that its power, when arriving at the legacy receiver as interference, does not exceed a certain prescribed threshold. Consequently, a class of optimization problems that seek to maximize the cognitive achievable rate subject to legacy received power constraints can be formulated and solved; see \cite{gastpar07:it} and subsequent works.

The aforementioned approach based upon regulating legacy received power essentially treats the interference power itself as an indicator of the interference level. This is consistent with the current FCC regulatory policy that measures the interference in terms of its power and calls the power level the ``interference temperature'' \cite{fcc}. A closer look at the cognitive transmission problem, however, suggests that it is not merely the power level itself, but the power spectral density (PSD), that dictates the achievable performance of a legacy receiver. This observation is the key idea underlying the development in our paper.

Starting from the idea of optimizing the PSD of cognitive signal (called ``spectrum shaping'' in this paper), we treat two representative cases: (1) where the legacy service does not encode its message and thus transmits an uncoded signal, and (2) where the legacy message is digitally coded. To keep the development more accessible while conveying the essential conclusions, in this paper we only consider a single pair of cognitive transceivers (the problem of spectrum sharing among multiple pairs of cognitive transceivers is left for future investigation).

Our main finding is that, surprisingly, through appropriately shaping the cognitive signal spectrum coupled with optimal filtering (for the case of uncoded legacy transmission) or decoding (for the case of coded legacy transmission), cognitive transmission can actually overcome the interference limitation, and can always transmit at its full available device power. This is in contrast to the aforementioned design principle based upon regulating legacy received power, because there the cognitive transmitter is forbidden to keep ramping its power level when hitting a certain upper limit, which is usually much smaller than the available device power especially when the distance between the cognitive transceivers is short, a typical scenario in proposed cognitive radio applications.

We present analytical development and numerical illustration for our results, in particular focusing on the logarithmic growth rate, {\it i.e.}, the prelog coefficient, of cognitive transmission in the high-power regime. As suggested by the previous discussion, this regime is the most relevant for applications, and also demonstrates the most of the advantages of the spectrum shaping approach. Briefly speaking, our results show that it is always possible to achieve logarithmic growth of cognitive transmission rates whenever the legacy channel is not fully loaded. The degree of legacy channel loading, nevertheless, affects the logarithmic growth rate, -- the heavier (lighter) the legacy channel is loaded, the larger (smaller) the prelog coefficient the cognitive transmission is allowed to achieve.

Before entering the detailed exposition of the paper, we briefly remark on the connection of our development to some apparently related work. As will be evidenced in this paper, the spectrum shaping approach typically leads to a subset of spectral bands that are ``turned on'' while leaving the remaining spectral bands ``off''. This appears to be analogous to frequency-division multiple access (FDMA), which has been shown in \cite{etkin07:jsac} to bear certain optimal properties in spectrum sharing among multiple cognitive systems for unlicensed bands ({\it i.e.}, without a legacy system). We note that, in our model, we are concerned with the coexistence between legacy and cognitive systems, and their sharing is not FDMA, because the legacy signal is assumed to occupy the entire spectrum. The ``on-off'' behavior of cognitive signal PSD is the more a consequence of non-convex optimization.

Another proposed approach with a strong information-theoretic flavor is based on the technique of ``dirty paper coding'' \cite{costa83:it}, as established in \cite{devroye06:it, jovicic06:isit}. A key assumption there is that the cognitive transmitter has access to the message (even not only the codebook) of the legacy transmitter. With this additional knowledge, the cognitive transmitter thus can adapt its signal using dirty paper coding so as to compensate for the excess interference it causes to the legacy receiver, and the overall effect is that the legacy transmission is not affected by the cognitive transmission. Similar to the spectrum sharing approach, this approach also achieves logarithmic growth of cognitive transmission rates, and furthermore does not reduce its prelog coefficient. As such, a distributed implementation of dirty paper coding has been instrumental for such results. Disregarding the attendant complexity and non-ideal effects in implementing such schemes in practice (see, {\it e.g.}, \cite{grover07:isit}), there remain a few more pressing concerns, such as scalability, privacy, and security -- thus at present, legacy message sharing is unlikely feasible due to technological and political constraints. In contrast, the spectrum sharing approach presented herein may be implemented using off-the-shelf technologies.

The remainder of the paper is organized as follows. The main body is divided into two parts, one (Part I) for the case of uncoded legacy transmission, and the other (Part II) for the coded case. Part I contains five sections. Section \ref{sec:model} introduces the basic system model, which is special as the legacy and cognitive receivers are collocated (whereas separately processed), -- an assumption adopted to minimize technical hurdles of analysis. Section \ref{sec:memoryless} gives the performance analysis when the legacy receiver ignores the inherent correlation structure in signals, {\it i.e.}, when taking an interference-temperature approach. The observation there is that the achievable cognitive rate is ultimately limited by the barrier posed by its interference to legacy users. The main section, Section \ref{sec:ncwk}, formulates the spectrum shaping problem, discusses its main properties, and gives necessary conditions for achieving its optimal solution. Special attention is paid to the on-off cognitive PSD, and the resulting prelog coefficient is analyzed. Section \ref{sec:case} presents two case studies to illustrate the analytical results, and numerical results show that the spectrum shaping approach leads to substantial rate gains compared with the interference-temperature approach. Lastly in this part, Section \ref{sec:general-topology} discusses the more general situation where there are multiple legacy receivers and the cognitive receiver is not collocated with them.

The treatment for the coded of coded legacy transmission is in Part II, which contains three sections. Section \ref{sec:model-coded} introduces the basic systems model, consisting of a pair of cognitive transceivers in the presence of a pair of legacy transceivers, with each transceiver being equipped with a single antenna. Section \ref{sec:prelog} studies various possible system behaviors in detail, and establishes that the cognitive rate prelog coefficient takes a particularly simple form, as one minus the normalized load of the legacy channel. Such a result thus evidently reveals the tradeoff between loading the legacy system and increasing the rate of cognitive transmission. Section \ref{sec:mimo} further extends the above result to a more general situation, where the cognitive transceivers are equipped with multiple antennas, and shows that increasing the number of antennas scales the prelog coefficient, thus compensating for the rate loss due to legacy transmission.

Finally, a section of concluding remarks concludes the paper, remarking on the system models used in the paper, and discussing some implementation issues.

\section*{Part I: Uncoded Legacy Transmission}

\section{Basic System Model: Collocated Receivers}
\label{sec:model}

In this section, we introduce a simplified channel model, in which the cognitive receiver and the legacy receiver are collocated such that they receive the common signal, as
\begin{eqnarray}
\label{eqn:Rx-signal-colloc}
\rvy_n = \sqrt{a} \rvs_n + \rvx_n + \rvn_n,
\end{eqnarray}
where the subscript $n$ denotes sampled discrete time. The legacy signal $\rvs_n$ is from a typically remote transmitter of some legacy service, for example, television or radio broadcasting, and is in uncoded analog form. The deterministic channel gain $a$ models the attenuation for the legacy signal at the receiver, -- it is assumed known at the receiver. The noise process $\rvn_n$ is comprised of thermal noise, unmodeled external random disturbance, and interference from other devices not considered in the system model. The cognitive transmitter sends digitally coded signals, with the resultant received signal, $\rvx_n$. A key observation we make in the paper is that the legacy channel gain $a$ typically has a large dynamic range (tens of dBs), thus leading to a large SNR margin that will be exploited by the cognitive user(s). In contrast, the cognitive transmitter is of interest only to its corresponding receiver, and we consequently normalize the cognitive signal's attenuation to unity, for simplicity and without loss of generality in the collocated receiver model.

We note that, although the model (\ref{eqn:Rx-signal-colloc}) describes the received signal at both a legacy receiver and a cognitive receiver, the two receivers have distinct goals. The legacy receiver simply wishes to reconstruct its legacy signal and has no knowledge of the cognitive user save its statistics (assumed to be passively identified via channel outputs). The cognitive receiver wishes to decode its message possibly exploiting statistical knowledge of the legacy signal and channel. Throughout the paper, we do not allow the cognitive transmitter to obtain the message of the legacy service provider, and deem such message-sharing as technologically and politically infeasible.

We further model the legacy signal $\rvs_n$ as a time-correlated stochastic process, taking continuous values on the complex plane. Specifically, we assume that $\rvs_n$ is a wide-sense stationary (WSS) zero-mean circular-symmetric complex Gaussian (ZMCSCG) process with PSD $\phi_\rvs(\omega)$, for $-\pi \leq \omega \leq \pi$, which is known as statistical knowledge at both the legacy receiver and the cognitive receiver. The noise $\rvn_n$ is also modeled as a WSS ZMCSCG process with a known PSD $\phi_\rvn(\omega)$.

\begin{figure}[t]
\centerline{\includegraphics[scale=0.45]{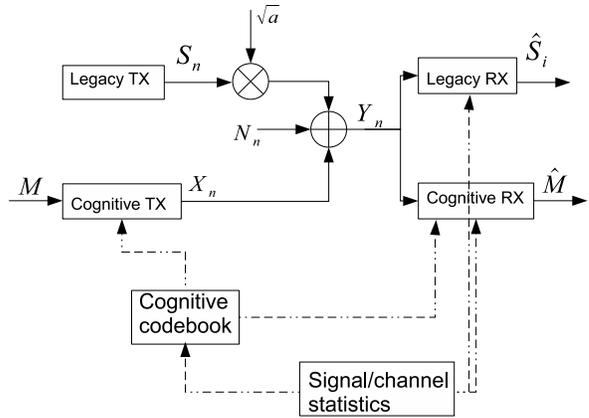}}
\caption{Channel model of cognitive radio with uncoded legacy service and collocated receivers.}
\label{fig:uncodedscheme}
\end{figure}

The legacy receiver does not possess any knowledge of the codebook of the cognitive signal $\rvx_n$, except certain statistics like the PSD, and thus can only estimate $\rvs_n$ by treating $(\rvx_n + \rvn_n)$ as some form of noise. Furthermore, the legacy service requires that the estimate of $\rvs_n$ satisfies a prescribed distortion constraint. We take the distortion measure as the MSE in the paper. Meanwhile, the cognitive receiver seeks to reliably decode its signal $\rvx_n$, with a vanishing message error probability for asymptotically large coding block length. Since the legacy signal $\rvs_n$ is uncoded, the best the cognitive receiver can do is to treat it as a time-correlated Gaussian noise, in addition to the actual system noise $\rvn_n$. Figure \ref{fig:uncodedscheme} illustrates the channel model.

Ignoring the legacy service's performance, the optimal input $\rvx_n$ for maximizing the cognitive transmission rate is a WSS ZMCSCG process, with its PSD $\phi_\rvx(\omega)$ chosen to solve a water-filling problem for (\ref{eqn:Rx-signal-colloc}). Given the legacy service and its distortion constraint, however, this water-filling PSD solution may no longer be optimal. Furthermore, if the legacy receiver takes into account the full statistics of $\rvx_n$ in estimating $\rvs_n$, then the optimal distribution of $\rvx_n$ may not even be Gaussian. In this paper, for tractability and practicality, we nevertheless restrict the legacy receiver to perform linear estimation and the cognitive transmitter to transmit WSS signal. Thus, estimation MSE depends only on the PSD of $\rvs_n, \rvx_n$, and $\rvn_n$ (see, {\it e.g.}, \cite{poor94:book}). Consequently, there is no loss in optimality to take $\rvx_n$ to be a WSS ZMCSCG process, and the problem reduces to designing its PSD $\phi_\rvx(\omega)$.

\section{An Interference-Temperature Perspective: Memoryless Estimation}
\label{sec:memoryless}

In this section, we consider a suboptimal estimation procedure, in which the legacy receiver performs a symbol-by-symbol minimum mean-squared error (MMSE) estimation of $\rvs_n$, without taking into account the temporal correlation structure of signals. Let us denote the variances of $\rvs_n, \rvx_n$ and $\rvn_n$ be $\sigma^2_\rvs$, $\sigma^2_\rvx$, and $\sigma^2_\rvn$, respectively. The MSE achieved by memoryless MMSE estimation of $\rvs_n$ is
\begin{eqnarray}
\mathrm{MSE}_0 = \frac{\sigma^2_\rvs (\sigma^2_\rvx + \sigma^2_\rvn)}{a \sigma^2_\rvs + \sigma^2_\rvx + \sigma^2_\rvn}.
\end{eqnarray}
Consequently, if the legacy receiver poses an MSE distortion constraint $D$ upon its estimation of $\rvs_n$, we have that the cognitive signal PSD should satisfy
\begin{eqnarray}
\label{eqn:memoryless-varx}
\sigma^2_\rvx = \frac{1}{2\pi} \int_{-\pi}^\pi \phi_\rvx(\omega) d\omega \leq \frac{\sigma^2_\rvs D}{\sigma^2_\rvs - D} a - \sigma^2_\rvn.
\end{eqnarray}

In the presence of an additional power constraint $P$ for $\rvx_n$, (\ref{eqn:memoryless-varx}) is simply another legacy-induced power constraint. Therefore the cognitive rate maximization problem is reduced to the following standard form,
\begin{eqnarray}
\label{eqn:memoryless-opt}
\max_{\phi_\rvx(\omega)}&& \frac{1}{2\pi} \int_{-\pi}^\pi \log\left(1 + \frac{\phi_\rvx(\omega)}{a \phi_\rvs(\omega) + \phi_\rvn(\omega)}\right) d\omega,\\
&&\mbox{s.t.}\quad \frac{1}{2\pi} \int_{-\pi}^\pi \phi_\rvx(\omega) d\omega \leq \overline{P}_0\nonumber\\
&&\quad\quad\quad\quad\quad:= \min\left\{P, \frac{\sigma^2_\rvs D}{\sigma^2_\rvs - D}a - \sigma^2_\rvn\right\}.
\end{eqnarray}
It readily follows that the standard water-filling strategy solves the problem, and the optimal solution is
\begin{eqnarray}
\phi_{\rvx, 0}(\omega) = \left[\frac{1}{\lambda} - a \phi_\rvs(\omega) - \phi_\rvn(\omega)\right]^+,
\end{eqnarray}
where $[\cdot]^+$ denotes $\max\{\cdot, 0\}$, and $\lambda > 0$ is chosen such that $\frac{1}{2\pi} \int_{-\pi}^\pi \phi_{\rvx, 0}(\omega) d\omega = \overline{P}_0$ is satisfied.

For memoryless MMSE, we emphasize that the achievable rate of the cognitive transmission is ultimately limited by the legacy channel gain $a$. For cognitive radio applications, the typically short-range channel for $\rvx_n$ makes the power constraint $P$ usually large. Due to the upper limit imposed by $a$, however, the maximum power of the cognitive transmission is severely regulated by $\overline{P}_0$, and thus the cognitive rate $R_{\mathrm{c}}$ behaves like $O(1)$ as $P \rightarrow \infty$, instead of $O(\log P)$ without the legacy MSE distortion constraint.

Finally, we note that, the cognitive rate maximization problem (\ref{eqn:memoryless-opt}) is feasible only under the conditions of
\begin{eqnarray}
\label{eqn:memoryless-cond1}
D &\leq& \sigma^2_\rvs\\
\label{eqn:memoryless-cond2}
\mbox{and}\quad D &\geq& \underline{D}_0 := \left(\frac{1}{\sigma^2_\rvs} + \frac{a}{\sigma^2_\rvn}\right)^{-1}.
\end{eqnarray}
The first condition (\ref{eqn:memoryless-cond1}) is obvious, since even without any channel output, the MSE distortion in estimating $\rvs_n$ is at most $\sigma^2_\rvs$. In other words, if $D \geq \sigma^2_\rvs$, the cognitive transmission is no longer constrained by the legacy service, and thus standard water-filling can be utilized with $\sigma^2_\rvx = P$. The second condition (\ref{eqn:memoryless-cond2}) poses a limit upon the minimum legacy channel gain $a$. That is, if the receiver is located such that $a$ is too small, then the distortion target $D$ cannot be satisfied even without any cognitive transmission, and the problem (\ref{eqn:memoryless-opt}) becomes infeasible.

\section{Optimal Legacy Receiver Processing: Non-Causal Wiener-Kolmogorov Filtering}
\label{sec:ncwk}

Given a sufficiently long block of channel outputs, the MMSE for estimating the legacy signal is achieved by the non-causal Wiener-Kolmogorov filter. Therefore as the observation interval grows long, the achievable MSE asymptotically approaches \cite{poor94:book}
\begin{eqnarray}
&&\mathrm{MSE}_\infty\nonumber\\
&=& \frac{1}{2\pi} \int_{-\pi}^\pi \frac{\phi_\rvs(\omega)[\phi_\rvx(\omega) + \phi_\rvn(\omega)]}{a \phi_\rvs(\omega) + \phi_\rvx(\omega) + \phi_\rvn(\omega)} d\omega\nonumber\\
&=& \sigma^2_\rvs - \frac{1}{2\pi}\int_{-\pi}^\pi \frac{a \phi_\rvs^2(\omega)}{\phi_\rvx(\omega) + a \phi_\rvs(\omega) + \phi_\rvn(\omega)} d\omega.
\end{eqnarray}
Thus for a prescribed MSE distortion constraint $D$, we can pose the cognitive rate maximization problem as follows.
\begin{eqnarray}
\label{eqn:ncwk-problem}
&&\max_{\phi_\rvx(\omega)} \frac{1}{2\pi} \int_{-\pi}^\pi \log\left(1 + \frac{\phi_\rvx(\omega)}{a \phi_\rvs(\omega) + \phi_\rvn(\omega)}\right) d\omega,\\
&&\mbox{s.t.}\nonumber\\
\label{eqn:ncwk-legacy-con}
&& \frac{1}{2\pi}\int_{-\pi}^\pi \frac{a \phi_\rvs^2(\omega)}{\phi_\rvx(\omega) + a \phi_\rvs(\omega) + \phi_\rvn(\omega)} d\omega \geq \sigma^2_\rvs - D\\
\label{eqn:ncwk-power}
&& \frac{1}{2\pi} \int_{-\pi}^\pi \phi_\rvx(\omega) d\omega \leq P.
\end{eqnarray}
That is, we maximize the information rate of the cognitive transmission, under the constraints that the estimation MSE of the legacy signal does not exceed a prescribed distortion $D$, and that the power of the cognitive signal does not exceed a maximally allowed value $P$.

\subsection{The Power Constraint is Always Active}
\label{subsec:active}

\begin{figure}[t]
\centerline{\includegraphics[scale=0.5]{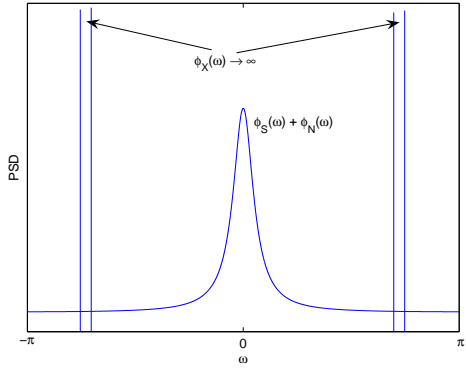}}
\caption{Illustration of $\phi_\rvx(\omega)$ that leads to divergent cognitive rates.}
\label{fig:unbounded}
\end{figure}

A surprising observation upon inspecting the cognitive rate maximization problem is that, if there is no power constraint (\ref{eqn:ncwk-power}) with $P < \infty$ and if the MSE constraint (\ref{eqn:ncwk-legacy-con}) leads to a strictly nonempty feasible set for $\phi_\rvx(\omega)$, then the cognitive rate in (\ref{eqn:ncwk-problem}) can be divergent, {\it i.e.}, infinitely large. This observation can be established as follows; see also Figure \ref{fig:unbounded}. If we let $\phi_\rvx(\omega) \rightarrow \infty$ over a pair of narrow intervals of frequencies $0 \leq \omega_l < |\omega| < \omega_u \leq \pi$, and let $\phi_\rvx(\omega) = 0$ elsewhere, then the resultant cognitive rate in (\ref{eqn:ncwk-problem}) diverges. On the other hand, the MSE constraint (\ref{eqn:ncwk-legacy-con}) now can be rewritten as
\begin{eqnarray*}
\frac{1}{2\pi} \int_{|\omega| \in [\omega_l, \omega_u]} \phi_\rvs(\omega) d\omega + \frac{1}{2\pi} \int_{|\omega| \notin [\omega_l, \omega_u]} \frac{\phi_\rvs(\omega) \phi_\rvn(\omega)}{a \phi_\rvs(\omega) + \phi_\rvn(\omega)} \leq D,
\end{eqnarray*}
which can be satisfied by choosing the volume of the interval $[\omega_l, \omega_u]$ sufficiently small but strictly positive, as long as we have
\begin{eqnarray}
\label{eqn:ncwk-feasible}
D > \underline{D} := \frac{1}{2\pi}\int_{-\pi}^\pi \frac{\phi_\rvs(\omega)\phi_\rvn(\omega)}{a \phi_\rvs(\omega) + \phi_\rvn(\omega)}d\omega.
\end{eqnarray}
In fact, a closer inspection of $\underline{D}$ reveals that it is exactly the MMSE achieved by non-causal Wiener-Kolmogorov filtering when we choose $\phi_\rvx(\omega) = 0$, $-\pi \leq \omega \leq \pi$. So (\ref{eqn:ncwk-feasible}) always holds true if the MSE constraint (\ref{eqn:ncwk-legacy-con}) leads to a strictly nonempty feasible set for $\phi_\rvx(\omega)$. Otherwise, when we have $D = \underline{D}$, the only feasible choice of $\phi_\rvx(\omega)$ is $\phi_\rvx(\omega) = 0$ for all $-\pi \leq \omega \leq \pi$ (possibly except for a subset of measure zero), and consequently the achievable cognitive rate vanishes.

\begin{figure}[t]
\centerline{\includegraphics[scale=0.5]{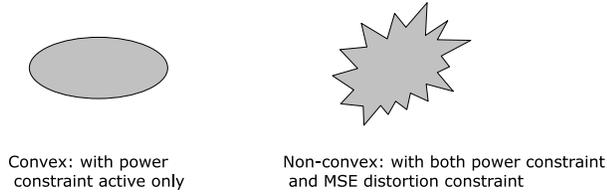}}
\caption{Illustration of the non-convexity of the feasible set of the cognitive PSD.}
\label{fig:feasible-set-illu}
\end{figure}

The observation that $\phi_\rvx(\omega)$ can lead to divergent cognitive rates while still satisfying the MSE distortion constraint immediately leads to two useful conclusions. First, the power constraint (\ref{eqn:ncwk-power}) must always hold with equality and thus is always active, for arbitrarily large $P$. In other words, there should not be any legacy-induced barrier that regulates the cognitive transmission power as in the memoryless estimation procedure described in Section \ref{sec:memoryless}. This conclusion is surprising, because as we commented in the introduction, without message-sharing, all the existing cognitive radio schemes are power-constrained so as not to exceed a certain externally imposed threshold for its interference to legacy receivers. Here, however, our observation implies that such a constraint is actually unnecessary and that the cognitive transmitter can indeed transmit with its full power. The second conclusion led by our observation is that, the cognitive rate maximization problem is generally non-convex, in the presence of the MSE distortion constraint. Intuitively, the feasible set of the cognitive PSD has many ``spikes'' that correspond to the peaky PSD as described in this subsection; see Figure \ref{fig:feasible-set-illu} for an illustration. The argument in our observation implies that it is often beneficial for the cognitive transmitter to operate close to those ``spikes''.

\subsection{General Necessary Condition for Optimal Cognitive PSD}
\label{subsec:necessary}

Depending upon whether the MSE distortion constraint (\ref{eqn:ncwk-legacy-con}) is active, the cognitive rate maximization problem exhibits two distinct cases, as elaborated in this subsection.

In \underline{the first case}, the power constraint $P$ is small, so that the water-filling solution of $\phi_\rvx(\omega)$ with respect to (\ref{eqn:ncwk-power}) does not violate the MSE distortion constraint (\ref{eqn:ncwk-legacy-con}). The water-filling solution,
\begin{eqnarray}
\label{eqn:case1-wf-sol}
&&\phi_{\rvx, \mathrm{wf}}(\omega) = \left[\frac{1}{\lambda} - a \phi_\rvs(\omega) - \phi_\rvn(\omega)\right]^+,\\
&&\mbox{s.t.}\;\; \frac{1}{2\pi}\int_{-\pi}^\pi \phi_{\rvx, \mathrm{wf}}(\omega)d\omega  = P,
\end{eqnarray}
is the optimal PSD for the cognitive transmission. For this case to arise, we need to have
\begin{eqnarray}
\label{eqn:case1con}
\frac{1}{2\pi}\int_{-\pi}^\pi \frac{a \phi_\rvs^2(\omega)}{\phi_{\rvx, \mathrm{wf}}(\omega) + a \phi_\rvs(\omega) + \phi_\rvn(\omega)} d\omega \geq \sigma_\rvs^2 - D,
\end{eqnarray}
where $\phi_{\rvx, \mathrm{wf}}$ is given by (\ref{eqn:case1-wf-sol}).

In \underline{the second case}, the inequality condition (\ref{eqn:case1con}) is violated, so it follows that both (\ref{eqn:ncwk-legacy-con}) and (\ref{eqn:ncwk-power}) constraints become active. The optimization problem is non-convex, since the feasible set of $\phi_\rvx(\omega)$ satisfying (\ref{eqn:ncwk-legacy-con}) is a non-convex set, as suggested by the previous subsection. In the following, we present a general necessary condition for optimality in this case.

\begin{prop}
\label{prop:necessary-general}
For the cognitive rate maximization problem (\ref{eqn:ncwk-problem}), in its second case where both the MSE distortion constraint and the power constraint are active, the optimal solution of the cognitive PSD $\phi_\rvx(\omega)$ should satisfy the following condition: for every $|\omega| \leq \pi$, either $\phi_\rvx(\omega)$ is a boundary point
\begin{eqnarray}
\label{eqn:boundary}
\phi_\rvx(\omega) = 0,
\end{eqnarray}
or $\phi_\rvx(\omega)$ is a locally maximum point
\begin{eqnarray}
\label{eqn:local-max}
\phi_\rvx(\omega) = \frac{\sqrt{1 + 4\lambda \mu a \phi_\rvs^2(\omega)} + 1}{-2\mu} - a \phi_\rvs(\omega) - \phi_\rvn(\omega),
\end{eqnarray}
for some $\mu < 0$ and $\lambda$ such that (a) the inequality
\begin{eqnarray}
\label{eqn:Delta-test}
1 + 4\lambda \mu a \phi_\rvs^2(\omega) \geq 0
\end{eqnarray}
holds, and (b) $\phi_\rvx(\omega)$ as given by (\ref{eqn:local-max}) is positive.
\end{prop}

Due to the lack of convexity in the optimization problem, in general, little can be said beyond Proposition \ref{prop:necessary-general} above. The solution of the cognitive rate maximization problem in the second case can in principle be obtained through an infinite-dimensional search procedure, outlined as follows. Consider an arbitrarily chosen measurable subset $\mathcal{I} \subseteq [0, \pi]$. Let $\phi_\rvx(\omega) = 0$ for every $|\omega| \notin \mathcal{I}$. Determine Lagrange multipliers $\mu < 0$ and $\lambda$ to generate $\phi_\rvx(\omega)$ according to (\ref{eqn:local-max}), for every $|\omega| \in \mathcal{I}$, such that both the MSE distortion constraint (\ref{eqn:ncwk-legacy-con}) and the power constraint (\ref{eqn:ncwk-power}) are satisfied with equality. Denote the resulting achievable rate calculated following (\ref{eqn:ncwk-problem}) by $R(\mathcal{I})$, then the maximum cognitive rate is $\sup R(\mathcal{I})$ with the supremum taken over all measurable subsets of $[0, \pi]$.

\subsection{A Suboptimal Solution: On-Off Cognitive PSD}
\label{subsec:onoff}

The fact that the power constraint (\ref{eqn:ncwk-power}) is always active implies that the cognitive transmitter can always transmit with its full power, without introducing excess interference to the legacy service, through appropriately tailoring its signaling PSD. As the power budget $P \rightarrow \infty$, the cognitive rate $R_\mathrm{c}$ behaves like $O(\log P)$, which is in fundamental contrast to the $O(1)$ behavior in the memoryless estimation scenario as described in Section \ref{sec:memoryless}.

A lower bound to the prelog coefficient for large $P$ can be obtained by considering a specific class of cognitive PSD, namely the on-off PSD for which $\phi_\rvx(\omega)$ equals either $0$ or some $\phi_0 > 0$. We optimize the prelog coefficient lower bound over all possible supporting sets of $\phi_0$, and the optimized lower bound is given by the following proposition.
\begin{prop}
\label{prop:onoff-prelog-lb}
For the cognitive rate maximization problem (\ref{eqn:ncwk-problem}), as the power budget $P \rightarrow \infty$, the on-off cognitive signal PSD that maximizes the prelog coefficient of the cognitive rate is determined by a parameter $\gamma > 0$ satisfying
\begin{eqnarray}
\label{eqn:onoff-prelog-gamma}
\frac{1}{2\pi} \int_{-\pi}^\pi \left[\frac{a \phi_\rvs^2(\omega)}{a \phi_\rvs(\omega) + \phi_\rvn(\omega)}\right]^-_\gamma = D - \underline{D},
\end{eqnarray}
where $[\cdot]^-_\gamma$ denotes $\min \{\cdot, \gamma\}$ and $\underline{D}$ is given by (\ref{eqn:ncwk-feasible}). The corresponding on-off cognitive signal PSD is then
\begin{eqnarray}
\phi_\rvx(\omega) = \phi_0 \cdot \mathrm{sgn}\left\{\left[\frac{a \phi_\rvs^2(\omega)}{a \phi_\rvs(\omega) + \phi_\rvn(\omega)}\right]^-_\gamma\right\},
\end{eqnarray}
where $\mathrm{sgn}\{\cdot\}$ denotes the sign of its operand, the on-level $\phi_0$ is chosen to satisfy the power constraint (\ref{eqn:ncwk-power}), and the prelog coefficient is the volume of the supporting set
\begin{eqnarray*}
\left\{\omega: \omega \in [0, \pi], \left[\frac{a \phi_\rvs^2(\omega)}{a \phi_\rvs(\omega) + \phi_\rvn(\omega)}\right]^-_\gamma > 0\right\}
\end{eqnarray*}
divided by $\pi$.
\end{prop}

As implied by (\ref{eqn:onoff-prelog-gamma}), the optimal choice of on-off PSD $\phi_\rvx(\omega)$ should ``fill'' the spectral bands in which the ``pre-emphasized'' PSD, $a \phi_\rvs^2(\omega)/[a \phi_\rvs(\omega) + \phi_\rvn(\omega)]$, is small, until the area (normalized by $2\pi$) under that PSD equals $(D - \underline{D})$, the gap between the prescribed MSE constraint and the MMSE without cognitive transmission.

\section{Case Studies}
\label{sec:case}

In order to gain insights from the analysis of memoryless estimation and non-causal Wiener-Kolmogorov estimation, we present two case studies in this section. For both case studies, the noise process $\rvn_n$ is memoryless with $\phi_\rvn(\omega) = \sigma_\rvn^2$, $-\pi \leq \omega \leq \pi$. The legacy signal $\rvs_n$ is taken as a memoryless process for the first case study, and as a first-order autoregressive (AR) process for the second. We note that, although singled out for its particular simplicity, the memoryless process is indeed a special first-order AR process with innovation rate unity.

\subsection{Memoryless Legacy Signal Process}
\label{subsec:memoryless}

In this case, we have $\phi_\rvs(\omega) = \sigma_\rvs^2$, $-\pi \leq \omega \leq \pi$. The analysis in Section \ref{sec:memoryless} for memoryless estimation immediately leads to
\begin{eqnarray}
\phi_{\rvx, 0}(\omega) = \overline{P}_0 = \min\left\{P, \frac{\sigma_\rvs^2 D}{\sigma_\rvs^2 - D}a - \sigma_\rvn^2\right\},\quad -\pi \leq \omega \leq \pi,
\end{eqnarray}
and
\begin{eqnarray}
R_\mathrm{c} = \log \left(1 + \frac{\overline{P}_0}{a \sigma_\rvs^2 + \sigma_\rvn^2}\right).
\end{eqnarray}

\begin{figure}[t]
\centerline{\includegraphics[scale=0.5]{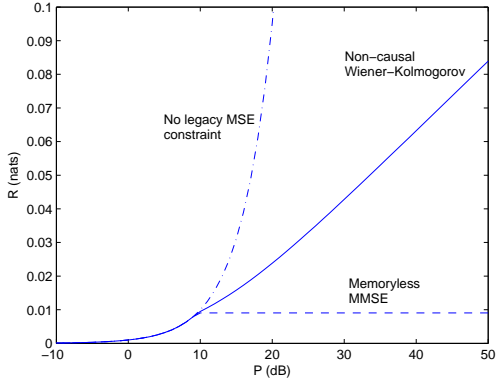}}
\caption{Illustration of the cognitive rates as a function of power $P$ for memoryless MMSE and non-causal Wiener-Kolmogorov filtering legacy receivers, in the case where the legacy signal process is memoryless. System parameters are $\sigma_\rvs^2 = 0$ dB, $\sigma_\rvn^2 = 0$ dB, $D = -20$ dB, and $a = 30$ dB.}
\label{fig:memoryless-psd-rates}
\end{figure}

For non-causal Wiener-Kolmogorov estimation, the first case for which water-filling with respect to the power constraint is optimal corresponds to $P \leq \frac{\sigma_\rvs^2 D}{\sigma_\rvs^2 - D}a - \sigma_\rvn^2$. So in the first case, the cognitive rate for non-causal Wiener-Kolmogorov estimation coincides with that for memoryless estimation. Beyond that threshold power, we enter the second case as elaborated in Section \ref{sec:ncwk}, and the spectrally flat $\phi_\rvs(\omega)$ and $\phi_\rvn(\omega)$ make the non-convex optimization problem easy to solve. In particular, it readily follows that the on-off cognitive PSD,
\begin{eqnarray*}
\phi_\rvx(\omega) = \phi_0 = \frac{a \sigma_\rvs^4 P}{(D - \underline{D})(a \sigma_\rvs^2 + \sigma_\rvn^2)} - a \sigma_\rvs^2 - \sigma_\rvn^2, \quad |\omega| \leq \frac{\pi P}{\phi_0},
\end{eqnarray*}
and $\phi_\rvx(\omega) = 0$ elsewhere, is an optimal solution. The maximum cognitive rate thus is
\begin{eqnarray}
R_\mathrm{c} = \frac{P}{\phi_0} \log \left(1 + \frac{\phi_0}{a \sigma_\rvs^2 + \sigma_\rvn^2}\right).
\end{eqnarray}
In Figure \ref{fig:memoryless-psd-rates}, we plot the cognitive rates as a function of power $P$ for memoryless MMSE and non-causal Wiener-Kolmogorov filtering legacy receivers. Comparing them, it is immediately seen that with the non-causal Wiener-Kolmogorov filter at the legacy receiver, the cognitive transmitter is able to achieve logarithmic growth in its rate; while with the memoryless MMSE estimation at the legacy receiver, the cognitive rate always dwells around small values due to the legacy-induced barrier upon the cognitive transmission power.

\begin{figure}[t]
\centerline{\includegraphics[scale=0.5]{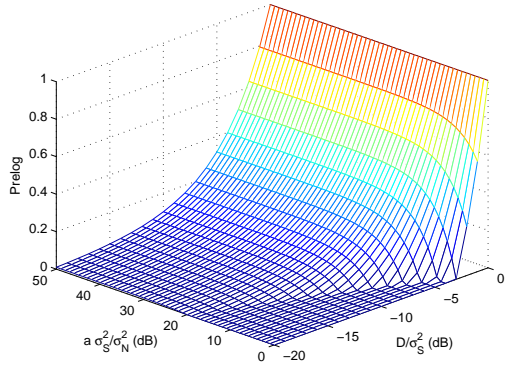}}
\caption{Illustration of the cognitive rate prelog coefficient as $P \rightarrow \infty$ versus both $D/\sigma_\rvs^2$ and $a \sigma_\rvs^2/\sigma_\rvn^2$, in the case where the legacy signal process is memoryless.}
\label{fig:prelog3dml}
\end{figure}

By noting that $\underline{D} = \frac{\sigma_\rvs^2 \sigma_\rvn^2}{a \sigma_\rvs^2 + \sigma_\rvn^2}$, the asymptotic prelog coefficient of $R_\mathrm{c}$ as $P\rightarrow\infty$ can further be rewritten as
\begin{eqnarray*}
\left(1 + \frac{\sigma_\rvn^2}{a \sigma_\rvs^2}\right)\cdot\frac{D}{\sigma_\rvs^2} - \frac{\sigma_\rvn^2}{a \sigma_\rvs^2}, \quad \mbox{or},\;\; \frac{D}{\sigma_\rvs^2} - \frac{\sigma_\rvn^2}{a \sigma_\rvs^2}\cdot \left(1 - \frac{D}{\sigma_\rvs^2}\right).
\end{eqnarray*}
We observe that, two factors affect the large-$P$ growth of $R_\mathrm{c}$. First, $D/\sigma_\rvs^2$ quantifies the quality requirement on estimating $\rvs_n$. The larger $D/\sigma_\rvs^2$ is, the coarser the legacy service quality we can tolerate, while the higher the cognitive rate we can achieve. Second, $a \sigma_\rvs^2/\sigma_\rvn^2$ quantifies the strength of the received legacy signal. For larger $a \sigma_\rvs^2/\sigma_\rvn^2$, the cognitive transmitter can exploit more spectral bands for its transmission and thus can achieve higher rates. In the limiting case, as $a \sigma_\rvs^2/\sigma_\rvn^2 \rightarrow \infty$, the prelog coefficient approaches $D/\sigma_\rvs^2$. The behavior of the prelog coefficient versus both $D/\sigma_\rvs^2$ and $a \sigma_\rvs^2/\sigma_\rvn^2$ is illustrated in the mesh plot of Figure \ref{fig:prelog3dml}. We also note that, if $a \sigma_\rvs^2/\sigma_\rvn^2$ is too small, then the cognitive rate maximization problem becomes infeasible, and the resulting prelog coefficient vanishes.

\subsection{First-Order AR Legacy Signal Process}
\label{subsec:1star}

In this case, $\rvs_n$ evolves as
\begin{eqnarray}
\rvs_{n + 1} = \sqrt{1 - \epsilon} \rvs_n + \sqrt{\epsilon} \rvw_n,
\end{eqnarray}
where the innovation rate $0 < \epsilon \leq 1$ characterizes the memory in process, and $\{\rvw_n\}_{n = 0}^\infty$ are a sequence of independent, identically distributed (i.i.d.) ZMCSCG random variables with variance normalized to $\sigma_\rvs^2$. Therefore the process $\rvs_n$ has PSD
\begin{eqnarray}
\phi_\rvs(\omega) = \frac{\epsilon\sigma_\rvs^2}{(2 - \epsilon) - 2\sqrt{1 - \epsilon} \cos\omega}, \quad -\pi \leq \omega \leq \pi.
\end{eqnarray}

\begin{figure}[t]
\centerline{\includegraphics[scale=0.5]{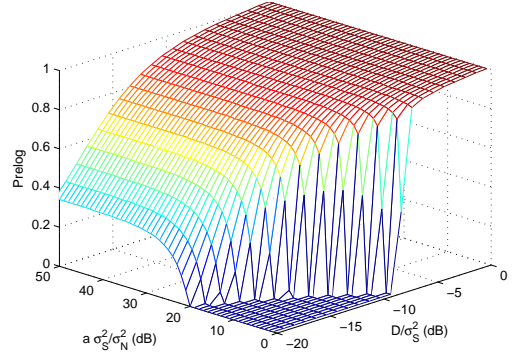}}
\caption{Illustration of the cognitive rate prelog coefficient as $P \rightarrow \infty$ versus both $D/\sigma_\rvs^2$ and $a \sigma_\rvs^2/\sigma_\rvn^2$, for $\epsilon = 0.1$ in the case where the legacy signal process is a first-order AR process.}
\label{fig:prelog3dar}
\end{figure}

We then examine the prelog coefficient for on-off cognitive PSD, following Section \ref{subsec:onoff}, and plot in Figure \ref{fig:prelog3dar} the mesh plot of the prelog coefficient versus both $D/\sigma^2_\rvs$ and $a \sigma^2_\rvs/\sigma^2_\rvn$, for $\epsilon = 0.1$, a moderate value of signal correlation. Compared with Figure \ref{fig:prelog3dml} which plots the case for memoryless legacy signal processes, we notice that legacy signal correlation significantly improves the achievable rate of cognitive transmission.

\begin{figure}[t]
\centerline{\includegraphics[scale=0.5]{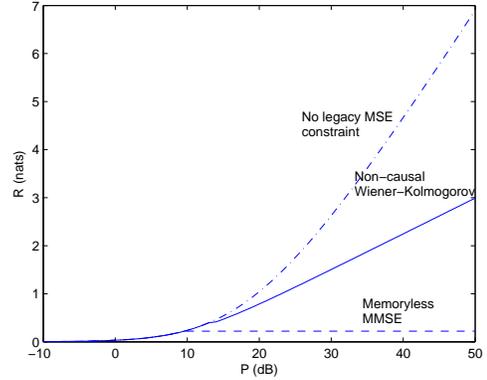}}
\caption{Illustration of the cognitive rates as a function of power $P$ for memoryless MMSE and non-causal Wiener-Kolmogorov filtering legacy receivers, in the case where the legacy signal process is a first-order AR process. System parameters are $\sigma_\rvs^2 = 0$ dB, $\sigma_\rvn^2 = 0$ dB, $D = -20$ dB, $a = 30$ dB, and $\epsilon = 0.1$.}
\label{fig:ar-psd-rates}
\end{figure}

In Figure \ref{fig:ar-psd-rates}, we plot the cognitive rates as a function of power $P$ for memoryless MMSE and non-causal Wiener-Kolmogorov filtering legacy receivers. Observations similar to that for Figure \ref{fig:memoryless-psd-rates} can be drawn by comparing the rates, and it can also be seen that the achievable rates substantially increase with increased legacy signal correlation.

\section{On-Off Cognitive PSD for General Legacy-Cognitive Topology}
\label{sec:general-topology}

\begin{figure}[t]
\centerline{\includegraphics[scale=0.5]{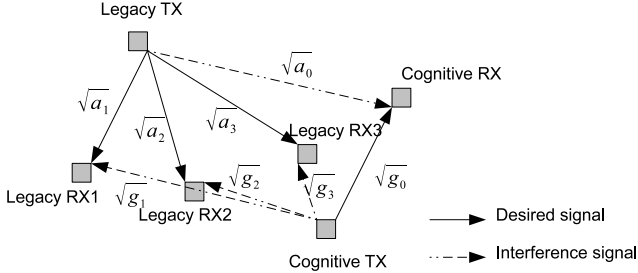}}
\caption{Illustration of a general cognitive radio system where there are multiple legacy receivers to be served and the cognitive receiver is not collocated with the legacy receivers.}
\label{fig:multilegacy}
\end{figure}

In this section, we briefly present the idea of how we may extend the insights from the basic system model of collocated receivers to the more general system model in which there are multiple legacy receivers not collocated with the cognitive receiver. This situation is as illustrated in Figure \ref{fig:multilegacy}. We assume that there are $K$ legacy receivers, each of which has channel gain $a_k$ from the legacy transmitter and channel gain $g_k$ from the cognitive transmitter, for $k = 1, \ldots, K$. We also denote the channel gain from the legacy transmitter to the cognitive receiver by $a_0$, and the channel gain from the cognitive transmitter to the cognitive receiver by $g_0$. For the $K$ legacy receivers, each has its own MSE distortion constraint, indexed as $D_k$, for $k = 1, \ldots, K$. More general scenarios of multiple legacy transmitters and multiple cognitive users for spectrum sharing are left for future investigation.

Analogously to the collocated-receiver system model, the cognitive rate maximization problem can be formulated. Using the same argument as in Section \ref{subsec:active}, we can again find that for sufficiently large $P$, the cognitive rate maximization problem is non-convex with the optimal cognitive PSD appropriately shaped. For general network topologies, however, an analytical characterization of the necessary condition for optimality does not appear to be immediately available. In the following, we therefore focus on the large-$P$ asymptotic analysis of the prelog coefficient, under the assumption of an on-off cognitive PSD.

Consider an on-off cognitive PSD $\phi_\rvx(\omega)$ satisfying $\phi_\rvx(\omega) = \phi_0$ if $|\omega| \in \mathcal{U}$ and $\phi_\rvx(\omega) = 0$ elsewhere, where $\mathcal{U}$ is a measurable subset of $[0, \pi]$. Following the same idea of Proposition \ref{prop:onoff-prelog-lb}, we can find that as $P \rightarrow \infty$, the prelog coefficient of the cognitive transmission rate is $|\mathcal{U}|/\pi$, and the support set $\mathcal{U}$ should satisfy the following set of inequalities:
\begin{eqnarray}
\label{ineq:multilegacy}
\frac{1}{\pi} \int_{\mathcal{U}} \frac{a_k \phi_\rvs^2(\omega)}{a_k \phi_\rvs(\omega) + \phi_{\rvn_k}(\omega)} d\omega \leq D_k - \underline{D}_k, \quad k = 1, \ldots, K,
\end{eqnarray}
where the MMSE without cognitive transmission $\underline{D}_k$ is
\begin{eqnarray}
\underline{D}_k = \frac{1}{2\pi}\int_{-\pi}^\pi \frac{\phi_\rvs(\omega) \phi_{\rvn_k}(\omega)}{a_k \phi_\rvs(\omega) + \phi_{\rvn_k}(\omega)} d\omega.
\end{eqnarray}
The optimization problem boils down to maximizing the volume $|\mathcal{U}|$ subject to the $K$ inequality constraints in (\ref{ineq:multilegacy}).

In general, the prelog maximization problem as formulated above is not amenable to efficient optimization algorithms. Here we only discuss one special case, to gain some practical insights.

{\it Low-Noise Limit:} In this asymptotic case, we let $\phi_{\rvn_k}(\omega) \rightarrow 0$ uniformly for all $-\pi \leq \omega \leq \pi$, and for all $k = 1, \ldots, K$. The practical scenario is that all the legacy receivers are close to the legacy transmitter so that the effect of noise can be neglected. The constraints (\ref{ineq:multilegacy}) now collapse into a single constraint
\begin{eqnarray}
\label{eqn:multilegacy-low-noise}
\int_\mathcal{U} \phi_\rvs(\omega) d\omega = \pi \min_{k = 1, \ldots, K} \left\{D_k - \sigma^2_{\rvn_k}/a_k\right\}.
\end{eqnarray}
So the optimal on-off cognitive PSD is obtained by choosing $\mathcal{U}$ to ``fill'' the spectral bands in which $\phi_\rvs(\omega)$ is small, until (\ref{eqn:multilegacy-low-noise}) is satisfied.

\section*{Part II: Coded Legacy Transmission}

\section{Basic System Model}
\label{sec:model-coded}

\begin{figure}[t]
\centerline{\includegraphics[scale=0.4]{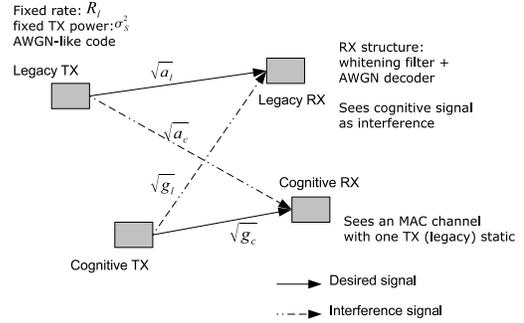}}
\caption{Illustration of the basic channel model for cognitive transmission in the presence of coded legacy transmission.}
\label{fig:scheme-coded-basic}
\end{figure}

In this section, we introduce the basic channel model, which consists of a pair of legacy transceivers and a pair of cognitive transceivers. All the transceivers are single-antenna and thus the signals are all scalar in the model. See Figure \ref{fig:scheme-coded-basic} for an illustration.
The legacy transmitter is static, since it exists before any potential cognitive transmission. We assume that the legacy transmitter employs a white ZMCSCG-like random codebook with a fixed rate $R_l$ and a fixed transmitted power $\sigma^2_\rvs$. For feasibility of cognitive transmission, we require that the rate $R_l$ is achievable at least without any cognitive transmission, {\it i.e.},
\begin{eqnarray}
\label{ineq:feasibility-basic-coded}
\mbox{Feasibility condition:}\quad \log\left(1 + \frac{a_l \sigma^2_\rvs}{\sigma^2_{\rvn_l}}\right) > R_l,
\end{eqnarray}
where $\sigma^2_{\rvn_l}$ is the power of the white ZMCSCG noise at the legacy receiver.

The cognitive transmitter is allowed to transmit a WSS ZMCSCG-like randomly coded message to the channel. The cognitive signal, upon arriving at the legacy receiver, is treated as a colored noise since the cognitive codebook is not known by the legacy transceivers, similar to the uncoded scenario treated in Part I of the paper. The legacy receiver, however, is assumed to be able to identify the statistics of the cognitive signal, and to perform the corresponding optimal decoding, which may be decomposed into a whitening filter stage and a decoding stage designed for white ZMCSCG noise. Denote the PSD of the cognitive signal by $\phi_\rvx(\omega)$, then we have the following two constraints:
\begin{eqnarray}
&&\mbox{Power constraint:}\quad \frac{1}{2\pi} \int_{-\pi}^\pi \phi_\rvx(\omega)d\omega \leq P\\
&&\mbox{Legacy rate constraint:}\nonumber\\
&&\quad \frac{1}{2\pi} \int_{-\pi}^\pi \log\left(1 + \frac{a_l \sigma^2_\rvs}{g_l \phi_\rvx(\omega) + \sigma^2_{\rvn_l}}\right) d\omega \geq R_l.
\end{eqnarray}

The cognitive receiver, however, does not treat the legacy signal as noise. This is because the cognitive service is built aware of the presence of legacy service, and the codebook of legacy transmission is broadcast in nature and thus publicly known at the cognitive receiver. The legacy transmitter, the cognitive transmitter, and the cognitive receiver hence form a multiple access (MAC) channel, whereas this MAC channel is not exactly the same as that classically treated in information theory ({\it e.g.}, \cite{cover91:book}), because here one of the transmitters (the legacy transmitter) is static and we can only design the other (the cognitive transmitter).

\section{Analysis and High-Power Regime Behavior}
\label{sec:prelog}

In this section, we analyze the operation of the cognitive transceivers.

{\it Case A:} The cognitive receiver cannot decode the legacy signal, even without any cognitive transmission. This case arises if the channel gain $a_c$ between the legacy transmitter and the cognitive receiver is small,
\begin{eqnarray}
\log\left(1 + \frac{a_c \sigma^2_\rvs}{\sigma^2_{\rvn_c}}\right) \leq R_l,
\end{eqnarray}
where $\sigma^2_{\rvn_c}$ is the power of the white ZMCSCG noise at the cognitive receiver. We can formulate the cognitive rate maximization problem under this case, as
\begin{eqnarray*}
\max_{\phi_\rvx(\omega)}&& \frac{1}{2\pi} \int_{-\pi}^\pi \log \left(1 + \frac{g_c \phi_\rvx(\omega)}{a_c \sigma^2_\rvs + \sigma^2_{\rvn_c}}\right)d\omega\\
&&\mbox{s.t.}\quad \frac{1}{2\pi} \int_{-\pi}^\pi \phi_\rvx(\omega)d\omega \leq P\\
&&\;\;\;\;\quad \frac{1}{2\pi} \int_{-\pi}^\pi \log\left(1 + \frac{a_l \sigma^2_\rvs}{g_l \phi_\rvx(\omega) + \sigma^2_{\rvn_l}}\right) d\omega \geq R_l.
\end{eqnarray*}

{\it Case B:} The cognitive receiver can decode the legacy signal. This case can be further refined into two sub-cases.

{\it Case B-1:} The cognitive receiver can utilize a successive decoding procedure, such that it first decodes the legacy signal treating the cognitive signal as noise, then decodes the cognitive signal after canceling the legacy signal. This situation is possible if
\begin{eqnarray}
\frac{1}{2\pi} \int_{-\pi}^\pi \log\left(1 + \frac{a_c \sigma^2_\rvs}{g_c \phi_\rvx(\omega) + \sigma^2_{\rvn_c}}\right)d\omega \geq R_l.
\end{eqnarray}
The resulting cognitive rate maximization problem becomes
\begin{eqnarray*}
\max_{\phi_\rvx(\omega)}&& \frac{1}{2\pi} \int_{-\pi}^\pi \log\left(1 + \frac{g_c \phi_\rvx(\omega)}{\sigma^2_{\rvn_c}}\right) d\omega\\
&&\mbox{s.t.}\quad \frac{1}{2\pi} \int_{-\pi}^\pi \phi_\rvx(\omega)d\omega \leq P\\
&&\;\;\;\;\quad \frac{1}{2\pi} \int_{-\pi}^\pi \log\left(1 + \frac{a_l \sigma^2_\rvs}{g_l \phi_\rvx(\omega) + \sigma^2_{\rvn_l}}\right) d\omega \geq R_l\\
&&\;\;\;\;\quad \frac{1}{2\pi} \int_{-\pi}^\pi \log\left(1 + \frac{a_c \sigma^2_\rvs}{g_c \phi_\rvx(\omega) + \sigma^2_{\rvn_c}}\right) d\omega \geq R_l.
\end{eqnarray*}

{\it Case B-2:} In this case, the cognitive receiver cannot reliably decode the legacy signal treating the cognitive signal as noise. Fortunately, we can utilize the idea of ``rate-splitting'' \cite{rimoldi96:it} to virtually split the cognitive signal into two parts, and accomplish successive decoding following the order of ``cognitive-legacy-cognitive''. Ignoring the incurred technical details here, we give the cognitive rate maximization problem formulation, as
\begin{eqnarray*}
\max_{\phi_\rvx(\omega)}&& \frac{1}{2\pi}\int_{-\pi}^\pi \log\left(1 + \frac{a_c \sigma^2_\rvs + g_c \phi_\rvx(\omega)}{\sigma^2_{\rvn_c}}\right) d\omega - R_l\\
&&\mbox{s.t.}\quad \frac{1}{2\pi} \int_{-\pi}^\pi \phi_\rvx(\omega)d\omega \leq P\\
&&\;\;\;\;\quad \frac{1}{2\pi} \int_{-\pi}^\pi \log\left(1 + \frac{a_l \sigma^2_\rvs}{g_l \phi_\rvx(\omega) + \sigma^2_{\rvn_l}}\right) d\omega \geq R_l\\
&&\;\;\;\;\quad \frac{1}{2\pi} \int_{-\pi}^\pi \log\left(1 + \frac{a_c \sigma^2_\rvs}{g_c \phi_\rvx(\omega) + \sigma^2_{\rvn_c}}\right) d\omega < R_l.
\end{eqnarray*}

\begin{figure}[t]
\centerline{\includegraphics[scale=0.5]{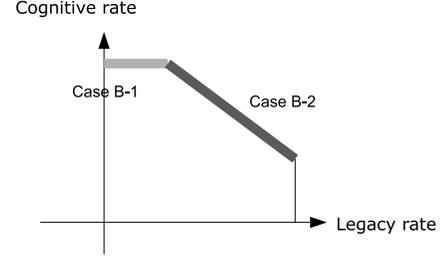}}
\caption{Illustration of the rate region of the MAC channel seen by the cognitive receiver. The horizontal segment of the rate region corresponds to Case B-1, and the slope segment (a.k.a. dominating face) corresponds to Case B-2.}
\label{fig:mac-illu}
\end{figure}

The two scenarios in Case B are illustrated in Figure \ref{fig:mac-illu}. Case B-1 occurs when the legacy rate $R_l$ is within the horizontal segment of the MAC rate region, so that direct successive coding is feasible. Case B-2 occurs when $R_l$ is within the slope segment (a.k.a. the dominating face of the MAC rate region), and then rate-splitting is necessary.

For both Case A and B, since we have assumed that both the legacy signal PSD and the noise PSD are flat, it straightforwardly follows from first-order optimality arguments analogous to that in Section \ref{subsec:necessary} that the optimal cognitive signal PSD is always an on-off PSD with the volume of the support spectral bands being optimized. Therefore we can readily obtain numerical values of the maximized cognitive rate under the various operational cases. We note that, in Case B, the cognitive transmitter can shape its signaling PSD $\phi_\rvx(\omega)$ to choose to operate either under Case B-1 or Case B-2. Depending upon the system parameters, either of the two sub-cases can lead to higher cognitive rates, so the globally optimal scheme should be obtained by comparing the resulting cognitive rates of the two sub-cases.

We are primarily interested in the regime where the cognitive transmission power budget $P$ is large. Similar to the prelog coefficient analysis in Section \ref{subsec:onoff} for the case of uncoded legacy transmission, here we evaluate the prelog coefficient of the cognitive rate as $P \rightarrow \infty$. We find that, although there are three possible operating cases (Cases A, B-1, and B-2), the maximum prelog coefficient is always unique as given by the following proposition.
\begin{prop}
\label{prop:coded-dof}
For cognitive transmission in the presence of coded legacy transmission, as the power budget $P \rightarrow \infty$, the optimal prelog coefficient of the cognitive rate is
\begin{eqnarray}
\Psi = 1 - \frac{R_l}{\log \left(1 + a_l \sigma^2_\rvs/\sigma^2_{\rvn_l}\right)}.
\end{eqnarray}
\end{prop}

Inspecting the expression of $\Psi$ in Proposition \ref{prop:coded-dof}, we can make two key observations. First, the growth rate of cognitive transmission only depends on parameters of the legacy channel. This observation is surprising, because one may initially think that the cognitive channel parameters, at least the interference (``cross-talk'') channel gains $a_c$ and $g_l$, would affect the asymptotic behavior of the cognitive rate. Second, noting that $\log\left(1 + a_l \sigma^2_\rvs/\sigma^2_{\rvn_l}\right)$ is nothing but the capacity of the legacy channel without cognitive transmission, the cognitive rate growth rate $\Psi$ is simply one minus the normalized load of the legacy channel. This observation is intuitively pleasing, because it quantitatively reveals the basic tradeoff between legacy and cognitive transmissions, -- the heavier (lighter) the legacy channel is loaded, the lower (higher) rate the cognitive transmission is allowed to achieve.

\begin{figure}[t]
\centerline{\includegraphics[scale=0.5]{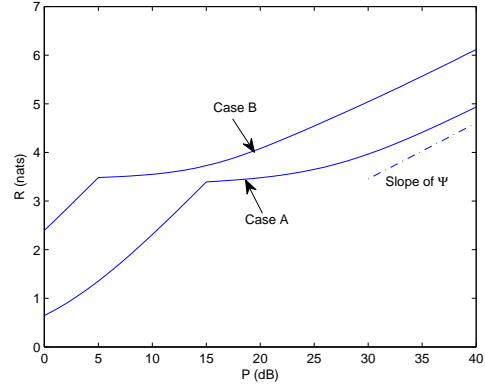}}
\caption{Cognitive rates as a function of the power constraint $P$, for both Case A and Case B with system parameters given in the example.}
\label{fig:rate-coded}
\end{figure}

{\it Example:} In this example, we plot cognitive rates in Figure \ref{fig:rate-coded}, for both Case A and Case B. The common system parameters are: $a_l = g_l = \sigma^2_{\rvn_l} = \sigma^2_{\rvn_c} = 0$ dB, $\sigma^2_\rvs = 30$ dB, $g_c = 10$ dB. In Case A, we take $a_c = -20$ dB, while in Case B $a_c = 0$ dB. The legacy rate $R_l$ is chosen to be half of the legacy channel capacity, hence the prelog coefficient is $\Psi = 0.5$, as indicated by the slope in the plot. It is evident that for both cases, the growth rates of cognitive transmission rate approach $\Psi$, as $P$ gets large.

\section{Extension to Cognitive Transceivers with Multiple Antennas}
\label{sec:mimo}

\begin{figure}[t]
\centerline{\includegraphics[scale=0.4]{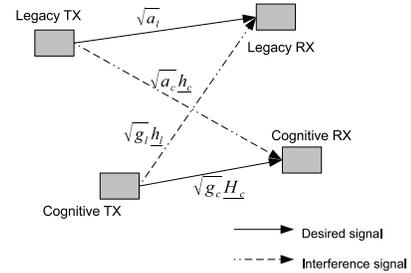}}
\caption{Illustration of the channel model for cognitive transmission in the presence of coded legacy transmission. The cognitive transceivers are equipped with multiple antennas while the legacy transceivers are scalar.}
\label{fig:scheme-coded-mimo}
\end{figure}

The analysis for the basic system model can be extended to the case where the cognitive transceivers are equipped with multiple antennas, as illustrated in Figure \ref{fig:scheme-coded-mimo}. We assume that the legacy transceivers are still scalar. The same as in the basic system model, the legacy transmitter employs static and white ZMCSCG-like random coding with rate $R_l$ and transmitted power $\sigma^2_\rvs$, and correspondingly the legacy receiver performs decoding after whitening the received signal.

To characterize the vector cognitive signal, we need to define its power spectrum density matrix; see, {\it e.g.}, \cite{kailath00:book} for a comprehensive introduction. In this section, we assume that the cognitive signal process $\underline{\rvx}_n$ is a WSS ZMCSCG vector process of dimension $N_t \times 1$, where $N_t$ is the number of transmit antennas. So the auto-correlation function of $\underline{\rvx}_n$ is
\begin{eqnarray}
K_{\underline{\rvx}}(k) = \mathbf{E}\left[\underline{\rvx}_{n + k} \underline{\rvx}_n^\dag\right],\quad k = \ldots, -1, 0, 1, \ldots,
\end{eqnarray}
where $[\cdot]^\dag$ denotes conjugate transpose of a complex-valued vector (or matrix, as in the following development). The PSD matrix of $\underline{\rvx}_n$ therefore is defined as the discrete-time Fourier transform of $K_{\underline{\rvx}}(k)$, as
\begin{eqnarray}
\phi_{\underline{\rvx}}(\omega) = \sum_{k} K_{\underline{\rvx}}(k) e^{-j\omega k}, \quad -\pi \leq \omega \leq \pi.
\end{eqnarray}

The vector cognitive signal collapses into a scalar signal at the legacy receiver, by the $N_t \times 1$ channel vector $\sqrt{g_l} \underline{h}_l$. The legacy receiver thus sees a channel as
\begin{eqnarray}
\rvy_{l, n} = \sqrt{a_l} \rvs_n + \sqrt{g_l} \underline{h}_l \underline{\rvx}_n + \rvn_{l, n}.
\end{eqnarray}
Hence the legacy rate constraint enforces the following inequality,
\begin{eqnarray}
\label{eqn:legacy-con-mimo}
\frac{1}{2\pi} \int_{-\pi}^\pi \log\left(1 + \frac{a_l \sigma^2_\rvs}{g_l \underline{h}_l \phi_{\underline{\rvx}}(\omega) \underline{h}_l^\dag + \sigma^2_{\rvn_l}}\right)d\omega \geq R_l.
\end{eqnarray}

The cognitive receiver sees a $N_r$-dimensional ($N_r$ denoting the number of receive antennas) vector MAC channel as
\begin{eqnarray}
\underline{\rvy}_{c, n} = \sqrt{g_c}\underline{H}_c \underline{\rvx}_n + \sqrt{a_c}\underline{h}_c \rvs_n + \underline{\rvn}_{c, n}.
\end{eqnarray}

In order to evaluate the achievable cognitive rate, we need a general formula that relates the entropy rate and the PSD matrix of a WSS ZMCSCG vector process. Such a formula was established in \cite{widom74:am}, and can be viewed as a generalization of the well-known theory for the asymptotic behavior of ordinary Toeplitz matrices and determinants (see, {\it e.g.}, \cite{gray06:notes}), which we have implicitly utilized throughout the development of the previous sections. Specialized to our problem, the basic result is that the entropy rate of a WSS ZMCSCG vector process is equal to
$1/(2\pi) \int_{-\pi}^\pi \log \det \phi(\omega) d\omega$, where $\phi(\omega)$ is the PSD matrix of the process.

Having introduced the above technical preparation, analysis of the operation of the cognitive transceivers closely follows that in Section \ref{sec:prelog} for scalar cognitive transceivers. Here we are again interested in the regime where the cognitive transmission power budget $P$ is large, and seek to identify the prelog coefficient of the cognitive rate. Note that the power constraint for vector cognitive transmission becomes
\begin{eqnarray}
\frac{1}{2\pi} \int_{-\pi}^\pi \mathrm{trace}[\phi_{\underline{\rvx}}(\omega)] d\omega \leq P.
\end{eqnarray}

Analogously to Proposition \ref{prop:coded-dof}, for vector cognitive transmission, the prelog coefficient is given by the following proposition.
\begin{prop}
\label{prop:coded-mimo-dof}
For cognitive transmission in the presence of coded legacy transmission, and with cognitive transceivers equipped with multiple antennas, as the power budget $P \rightarrow \infty$, the optimal prelog coefficient of the cognitive rate is
\begin{eqnarray}
\Psi = \left[1 - \frac{R_l}{\log \left(1 + a_l \sigma^2_\rvs/\sigma^2_{\rvn_l}\right)}\right] \cdot \mathrm{rank}\left[\underline{H}_c\right].
\end{eqnarray}
\end{prop}

Inspecting the expression of $\Psi$ in Proposition \ref{prop:coded-mimo-dof}, we notice that it is the product of two terms. The first term is nothing but $\Psi$ in Proposition \ref{prop:coded-dof} for scalar cognitive transmission, and thus the two key observations made therein apply. For the second term of the rank of the cognitive channel matrix $\underline{H}_c$, we can make two additional observations. First, since $\underline{H}_c$ is full-rank with high probability in a propagation environment of rich scattering, employing multiple antennas at cognitive transceivers is beneficial in that it scales the growth rate of the cognitive transmission, thus offsetting the loss due to loading the legacy channel. Second, $\Psi$ is independent of the spatial beam-forming used by the cognitive transmitter. In fact, as can be seen from the proof of Proposition \ref{prop:coded-mimo-dof}, the growth rate of $\Psi$ can be always achieved, as long as we choose the ``on'' PSD matrix $\underline{\phi}_0$ to satisfy $\mathrm{rank}[\underline{H}_c \underline{\phi}_0 \underline{H}_c^\dag] = \mathrm{rank}[\underline{H}_c]$, -- easily fulfilled by letting $\underline{\phi}_0 = I_{N_t \times N_t}$, for example. However, we note that appropriately shaping $\underline{\phi}_0$ may lead to additional rate gain in the regime of finite cognitive power budget, and this is a potential research topic for future investigation.

\section*{Concluding Remarks}

\subsection*{Remarks on System Models}

A few remarks regarding the assumptions of the system models used in the paper follow here.

In both cases (uncoded and coded legacy transmission) studied in the paper, the channel models are discrete-time. There are some delicate issues in such a modeling strategy, related to sampling and time synchronization. For the case of uncoded legacy transmission, the baseband legacy signal for transmission is generally not pulse-shaped and can be viewed as a continuous-time WSS ZMCSCG process. Therefore, to obtain the discrete-time channel model at the (legacy and cognitive) receivers, we simply need to sample the received signal in synchronization with the timing epochs of the digitally coded and pulse-shaped cognitive signal. Here, an implicit requirement is that the pulse width of the cognitive signal exactly coincides with the required sampling period of the legacy signal for estimation; in other words, the signaling bandwidth of the cognitive transmission is set at least equal to the bandwidth of the legacy signal, -- and the sampling period should always be no less than the signal bandwidth for alias-free estimation.

For the case of coded legacy transmission, both the legacy signal and the cognitive signal are pulse-shaped. The discrete-time channel models therefore imply that the two sets of transceivers are time-synchronized at the symbol level. For cognitive transceivers, synchronization may be achieved by locking into the timing epochs of the legacy signal, after an initial period of tracking, which is possible because after all we have assumed that the cognitive transmission system has knowledge of the legacy codebook and signaling schemes. If the cognitive transceivers do not make any effort in keeping synchronization with the legacy transceivers, then both (legacy and cognitive) receivers experience inter-symbol interference (ISI), and our analysis should be modified accordingly to account for the effect of ISI, -- a research topic for future investigation.

Our treatment of the case of uncoded legacy transmission only applies when the legacy signal is linearly modulated such that the message to be estimated is proportional to the actual signal transmitted. Examples of such systems include amplitude-modulation (AM) broadcast radio, analog television (NTSC, for example), and so on; but do not include nonlinearly modulated systems like frequency-modulation (FM) broadcast radio or wireless microphone (typically using FM). Analysis for those systems can be technically challenging due to the lack of convenient tools for quantifying the achievable MSE, especially in the presence of correlated cognitive signal.

\subsection*{Implementation Issues}

The envisioned implementation of the ideas studied in the paper can be an overlay cognitive system built in the presence of, instead of in the absence of, legacy transmission. As revealed by our analysis, the most benefits are obtained when the gap between the target and the achievable communication quality (in terms of estimation MSE or information rates) of the legacy service is large, {\it i.e.}, when the legacy channel is lightly loaded. With spectrum shaping, we can achieve logarithmic growth of cognitive transmission rates, even without knowledge of the legacy message at the cognitive transmitter. Certain modifications need to be implemented at the legacy receiver, in order to adapt to the statistics induced by the cognitive spectrum shaping. However, we note that such modifications do not demand excess increase in receiver complexity, and may be implemented for a dynamic environment through adaptive MMSE filtering or noise whitening.

Some more specific discussion about the implementation issues follows here.

\subsubsection*{Legacy Reception with Finite Buffering}
In the analysis of the case of uncoded legacy transmission, the optimal estimator at the legacy receiver is a non-causal Wiener-Kolmogorov filter, which, in principle, requires observations from the infinite future for processing. In implementation, we approximate the non-causal Wiener-Kolmogorov filtering operation by introducing a large, yet finite, block of delay. An interesting question then is how large the delay should be chosen, such that the estimation performance is satisfactory while the required buffer size and latency is tolerable. If we consider the extreme case, where we do not introduce any delay such that the legacy estimator is a pure predictor, then a related problem is how to optimize the cognitive rate by spectrum shaping. For causal estimation, the MSE formula is highly complex (see, {\it e.g.}, \cite{poor94:book} \cite{kailath00:book}), so this problem is non-trivial. An interesting question there is whether the cognitive transmission rate still logarithmically grows with the power budget.

\subsubsection*{Multicarrier Implementation}
Spectrum shaping implies that the coded symbols of cognitive signal are correlated as indicated by the PSD. A direct implementation for introducing the correlation is transmit side pre-coding. On the other hand, due to the increasing popularity of multicarrier systems like orthogonal frequency-division multiplexing (OFDM), it is useful to examine how spectrum shaping can be implemented through multicarrier modulation. In fact, since spectrum shaping manipulates the signal power across spectral bands, it is particularly suitable for multicarrier implementation. In doing so, the cognitive transmitter simply divides its whole bandwidth into multiple narrow bins, ensuring that the specified cognitive signal PSD does not rapidly change within each narrow bin. There is one caution to be exercised for coded legacy transceivers if they also employ multicarrier modulation. That is, the legacy message should be coded in a way such that the resulting signal is robust against the effect that some subcarriers are essentially ``erased'' due to the existence of ``on'' cognitive PSD in those subcarriers. Alternatively, the cognitive transceivers may spread out their interference to the legacy receiver, by pseudo-randomly changing the cognitive PSD such that the interfered subcarriers vary with time and no single subcarrier suffers from constant interference.

\section*{Acknowledgment}
The support of NSF OCE-0520324 is gratefully acknowledged.

\bibliographystyle{ieee}
\bibliography{./dyspan}

\begin{thebibliography}{1}

\bibitem{mitola00:phd}
J. Mitola,
\newblock \textit{Cognitive Radio: An Integrated Agent Architecture for Software Defined Radio},
\newblock Doctor of Technology, Royal Inst. Technol. (KTH), Stockholm, Sweden, 2000.

\bibitem{haykin05:jsac}
S. Haykin,
\newblock ``Cognitive Radio: Brain-Empowered Wireless Communications,''
\newblock \textit{IEEE J. Select. Areas Commun.}, Vol. 23, No. 2, pp. 201--220, Feb. 2005.

\bibitem{sahai06:report}
A. Sahai, N. Hoven, S.~M. Mishra, and R. Tandra,
\newblock ``Fundamental Tradeoffs in Robust Spectrum Sensing for Opportunistic Frequency Reuse,''
\newblock Technical Report, University of California, Berkeley, Mar. 2006.

\bibitem{dyspan:proc}
\textit{Proc. IEEE Symp. New Frontiers in Dynamic Spectrum Access Networks (DySPAN)},
\newblock 2005--.

\bibitem{jsac07:special}
Special Issue on ``Adaptive, Spectrum Agile and Cognitive Wireless Networks,''
\newblock \textit{IEEE J. Select. Areas Commun.}, Vol. 25, No. 3, Apr. 2007.

\bibitem{jsac08:special}
Special Issue on ``Cognitive Radio: Theory and Application,''
\newblock \textit{IEEE J. Select. Areas Commun.}, Vol. 26, No. 1, Jan. 2008.

\bibitem{gastpar07:it}
M. Gastpar,
\newblock ``On Capacity Under Receive and Spatial Spectrum-Sharing Constraints,''
\newblock \textit{IEEE Trans. Inform. Theory}, Vol. 53, No. 2, pp. 471--487, Feb. 2007.

\bibitem{fcc}
Spectrum Policy Task Force Reports, FCC.

\bibitem{etkin07:jsac}
R. Etkin, A. Parekh, and D. Tse,
\newblock ``Spectrum Sharing for Unlicensed Bands,''
\newblock \textit{IEEE J. Select. Areas Commun.}, Vol. 25, No. 3, Apr. 2007.

\bibitem{costa83:it}
M.~H.~M. Costa,
\newblock ``Writing on Dirty Paper,''
\newblock \textit{IEEE Trans. Inform. Theory}, Vol. 29, No. 3, pp. 439--441, May 1983.

\bibitem{devroye06:it}
N. Devroye, P. Mitran, and V. Tarokh,
\newblock ``Achievable Rates in Cognitive Radio,''
\newblock \textit{IEEE Trans. Inform. Theory}, Vol. 52, No. 5, pp. 1813--1827, May 2006.

\bibitem{jovicic06:isit}
A. Jovicic and P. Viswanath,
\newblock ``Cognitive Radio: An Information-Theoretic Perspective'',
\newblock in \textit{Proc. Int. Symp. Inform. Theory}, Seattle, WA, Jun. 2006.

\bibitem{grover07:isit}
P. Grover and A. Sahai,
\newblock ``Writing on Rayleigh Faded Dirt: A Computable Upper Bound to the Outage Capacity,''
\newblock in \textit{Proc. Int. Symp. Inform. Theory}, Nice, France, Jun. 2007.

\bibitem{poor94:book}
H.~V. Poor,
\newblock \textit{An Introduction to Signal Detection and Estimation},
\newblock Springer-Verlag, 2nd ed., 1994.

\bibitem{cover91:book}
T.~M. Cover and J.~A. Thomas,
\newblock \textit{Elements of Information Theory},
\newblock Wiley, 1991.

\bibitem{rimoldi96:it}
B. Rimoldi and R. Urbanke,
\newblock ``A Rate-Splitting Approach to the Gaussian Multiple-Access Channel,''
\newblock \textit{IEEE Trans. Inform. Theory}, vol.~42, no.~2, pp.~364--375,	Mar. 1996.

\bibitem{kailath00:book}
T. Kailath, A.~H. Sayeed, and B. Hassibi,
\newblock \textit{Linear Estimation},
\newblock Prentice-Hall, Upper Saddle River, NJ, 2000.

\bibitem{widom74:am}
H. Widom,
\newblock ``Asymptotic Behavior of Block Toeplitz Matrices and Determinants,''
\newblock \textit{Advances in Math.}, Vol. 13, pp. 284--322, 1974.

\bibitem{gray06:notes}
R.~M. Gray,
\newblock \textit{Toeplitz and Circulant Matrices: A Review}
\newblock NOW Publishers, 2006.

\end{thebibliography}

\end{document}